\documentstyle[twocolumn,aps]{revtex}
\begin{document}
\title{Multielectron dissociative ionization of molecules
       by intense laser radiation}
\author{Miros{\l}aw Brewczyk}
\address{Filia Uniwersytetu Warszawskiego,\\
 ul. Lipowa 41, 15-424 Bia{\l}ystok, Poland}
\author{Kazimierz Rz\c{a}\.zewski}
\address{Centrum Fizyki Teoretycznej PAN and College of Science,\\
  Al. Lotnik\'ow 32/46, 02-668 Warsaw, Poland}
\author{Charles W. Clark}
\address{Electron and Optical Physics Division, Physics Laboratory\\
National Institute of Standards and Technology, Technology Administration,\\ U.S. Department of Commerce, Gaithersburg, MD 20899}

\maketitle

\begin{abstract}
We solve the hydrodynamic-ballistic equations of motion for a
one-dimensional time-dependent Thomas-Fermi
model of Cl$_2$ exposed to an intense subpicosecond
laser field, and observe simultaneous multielectron 
ionization and molecular dissociation.  The fragment 
kinetic energy defect with respect to the simple Coulomb 
explosion picture is found to originate in ejected-electron
screening of the escaping fragments; its magnitude agrees 
with that observed in recent experiments.
\end{abstract}

\draft

Molecules exposed to intense ($I > 10^{13}$ W cm$^{-2}$) 
subpicosecond laser pulses undergo multiple ionization 
accompanied by dissociation, a process denoted by
multielectron dissociative ionization (MEDI)
~\cite{experiment,French,Froggy}.
The simplest picture of MEDI is
that of the "Coulomb explosion," in which the laser field
quickly strips a number of electrons from the molecule,
and the charged ionic fragments then dissociate under the
influence of the repulsive Coulomb potential.  In this
picture, the total kinetic energy of the fragment pair
with charges $q_1e$, $q_2e$ should be equal to
$T = q_1q_2e^2/(4\pi\epsilon_0R_e)$ (SI units),
where $q_1$ and $q_2$ are integers, $e$ is the elementary
charge, and $R_e$ is the equilibrium internuclear
separation of the neutral molecule.

In experiments, however, it is found that the fragment kinetic
energies are consistently lower than those predicted
by the Coulomb explosion model, by an amount
called the kinetic energy defect, $\Delta$.
This appears to be true for all observed fragmentation channels,
and does not depend much on
other experimental parameters such as laser wavelength
and pulse duration.  The fragment energies are
instead consistent with a Coulomb explosion at a value 
of $R$ some 20-50\% greater than $R_e$.
However, since the characteristic time scale for
molecular internuclear motion (a vibrational
period) is long compared to the time scale on which
strong-field multiple ionization occurs, it is
difficult to see how a simple Coulomb explosion
picture could be applicable.

This Letter puts forth a simple explanation
of these MEDI observations, and supports it by numerical
calculations applied to a model molecule.  A Coulomb explosion
is initiated at $R_e$, when the most loosely-bound
electrons are stripped from the molecule, but the
ejected electron cloud does not expand rapidly.
Ionization continues until
only a tightly-bound electronic core remains, and is
then shut off by the strong internal molecular field
(although for very intense radiation there may
be some subsequent ionization of the separated
fragments, i.e. post-dissociation ionization (PDI)).
Most electron stripping thus occurs 
near $R \approx R_e$, so one sees a
brief interval of acceleration of the
ionic fragments by increasing mutual Coulomb
repulsion.  However, when the ionization shuts
off, a countervailing {\it decelerating} tendency
becomes apparent:  as the fragments separate
through the ejected electron cloud, more electronic
charge is encompassed by the increasing internuclear
separation $R(t)$, and the resultant screening slows 
the fragments.  This post-explosion screening gives a 
value of $\Delta$ consistent with that observed in 
experiments.  Charge-symmetric fragmentation is confirmed
\cite{nitrogen} to be a major channel of dissociation.
These conclusions differ from
those of other models that have been
proposed recently ~\cite{French,nitrogen,theory,Ivanov-Corkum}.

We have done calculations to illustrate this
picture, by solving the time-dependent equations
of motion for the nuclei and
electrons of a model Cl$_2$ molecule in a strong
laser field.  All electrons are
treated explicitly, via time-dependent 
density functional theory.  We
know of no previous treatment of molecules
in strong fields that deals explicitly with more than
one electron.  Our approach is made as follows.

In strong-field laser irradiation of diatomic molecules,
it is believed \cite{Ivanov-Corkum}
that the molecular axis is quickly aligned with the polarization
vector of the radiation field, so that
the ionization process can be modelled with the
molecular axis parallel to the electric field.
To simplify the numerical calculations,
we make the additional approximation of
confining all particles in the system to
move in one dimension, defined by this axis.
This approximation has been used to
model strong-field interaction with one-and two-electron atoms
\cite{Eberly} and with the H$_2^+$ molecule \cite{KMS},
and it has replicated much of the essential physics of
three-dimensional systems.
We utilize time-dependent
density functional theory to deal directly with the electron
density vs. the many-electron Schr\"odinger
wavefunction.  We have previously applied time-dependent
density functional theory to treat three-dimensional
atoms in strong radiation fields \cite{PRA}, and obtained results
for multiple electron ionization that agree well
with experiments.

The electron density of the molecule
is described as a fluid of mass density
$\rho(x,t)$, with an associated velocity field $v(x,t)$,
which obey the equations of motion:

\begin{eqnarray}
\frac{\partial \rho}{\partial t} + \frac{\partial}{\partial x}\:
\left(\rho v \right) & = &  0  \nonumber  \\
\frac{\partial v}{\partial t} + v \frac{\partial v}
{\partial x} & = & - \frac{1}{\rho} \frac{\partial}{\partial x}
P + \frac{e}{m} \frac{\partial}{\partial x} \Phi
\label{hydr}
\end{eqnarray}

The first of the eqs. (\ref{hydr}) is the usual continuity equation.
The second is the classical equation
of motion for an infinitesimal element of fluid subject to forces
due to the gradients of an electrostatic
potential $\Phi$ and a pressure $P = -(\partial U / \partial x)_{s}$,
where $U$ is the internal energy density of the fluid.
The effects of quantum mechanics in this
equation must be expressed
by a constitutive relation between $P$ and $\rho$.
Such a relation was derived in a semiclassical approximation by
Thomas and Fermi (TF) \cite{TF}; its variants have 
been used to investigate the ground state properties
of atoms \cite{Spruch,Morgan} and weak-field atomic radiative response
\cite{Ball}. The TF model treats the
electrons as a Fermi gas at temperature $T= 0$, and
determines the energy of the gas by filling
the available phase-space volume, subject to
the Pauli exclusion principle.  This gives
$U$ of the three-dimensional
electron gas
as a local function of the density, of the
form $U({\bf r}) \sim \rho({\bf r})^{5/3}$.

We have sought a corresponding relationship applicable to the
one-dimensional many-electron system.  The Hohenberg-Kohn theorem
\cite{HK} ensures that the
energy of a stationary one-dimensional system is
a functional of the electron density, but
the semiclassical TF
arguments may not provide a good approximation
to the true energy functional. 
Experimentation with possible forms of the
functional leads us to propose the following
expression for the energy density of a one-dimensional electron
gas, $U(x) = A n^{2}(x)$,
where $n(x)$ is the electron density function $(n(x)=\rho(x)/m)$,
and $A$ is a universal constant.  The value of
$A$ is determined by requiring that it give
reasonable physical properties (binding energies, etc.)
of the model systems.  Such an expression is also obtained
by applying the semiclassical method in
{\it two} dimensions \cite{2D}.

In treating Coulomb interactions in one dimension,
it is necessary to eliminate the singularity in the potential
at $x = 0$.  We follow a standard procedure \cite{Eberly,KMS},
and write the electrostatic potential $\Phi$ of our model
diatomic molecule as
\begin{eqnarray}
\Phi & = & \frac{Z e}{(b^{2}+(x-x_1(t))^{2})^{1/2}}\; + \:
              \frac{Z e}{(b^{2}+(x-x_2(t))^{2})^{1/2}} \nonumber \\
 & &  - \; e \int_{-\infty}^{\infty} \frac {n(x^{\:\prime},t)}
    {(c^{2} + (x-x^{\:\prime})^{2})^{1/2}} \:\: dx^{\:\prime} \; .
\label{pot}
\end{eqnarray}
Here $Z$ is the atomic number and $x_1(t),x_2(t)$ are positions of the
two nuclei.
The parameters $b, c$ serve to smooth the Coulomb
interaction at points of coalescence.

By solving the static case of eq. \ref{hydr}, we obtain the
linear relationship, $n(x)=e \Phi (x)/(2 A)$,
which provides an integral equation for the
electrostatic potential of the time-independent
solution \cite{BRC}. Its linearity allows us to construct the solution
for a many-atom system by linear combination of appropriately
shifted single-atom solutions. It is easy to show that the
parameters $A$, $b$, and $c$ are not independent \cite{BRC}:
$A = e^{2} \ln (c/b)$. Thus we always have $c > b$. Furthermore,
our numerical calculations
of values of $R_e$ and $D$ for various values of $Z$ indicate that
the value of $A$ must be essentially independent of
$Z$, and that $A = 0.3$ a.u. [in the usual atomic units (a.u.)
in which the numerical values of the electron
mass $m$, the elementary charge $e$, and reduced Planck's
constant $\hbar$ are
equal to unity]. If $A$ lies outside a narrow band
of values around 0.3 a.u., one obtains unreasonable
values of $R_e$ and $D$.

We have applied this model to describe the
Cl$_2$ ($Z=17$) molecule in a strong radiation field.  By solving the
static equation as a function of internuclear distance,
we find the Born-Oppenheimer potential curves for
Cl$_2$ and Cl$_2^{+2}$ displayed in Fig. 1.  These results
were obtained with the parameter set
($A=0.3$ a.u., $b=1.22$ a.u., $c=1.65$ a.u.)
which yields $R_e = 3.8$ a.u. and
$D = 0.11$ a.u., in agreement
with the experimental values \cite{Herzberg} of $R_e = 3.76$ a.u. and
$D = 0.091$ a.u. respectively.

We treat the time-dependent system by adding the 
electric dipole interaction with the radiation field to (\ref{pot}),
and solving
the hydrodynamic equations for the electronic density simultaneously
with the classical equations of motion for the positions of the nuclei.
We treat the equations (\ref{hydr})
as an initial value problem, with the density $\rho(x,t=0)$ given by
the static solution, and the nuclei
initially at rest and separated by $R_e$.
The first check of
the time-dependent method is to verify that it reproduces
molecular vibrations when no external field is present.
That it does is shown in the inset of Fig. \ref{BOcurves},
which displays the free oscillation of the internuclear
distance in the absence of a radiation field.  The period of
oscillation of 34 fs is shorter than the experimental
value \cite{Herzberg} of 59 fs, so our model overestimates
the stiffness of the molecular potential at $R_e$.

We now discuss the effects of an intense subpicosecond
laser pulse on this system.  The laser frequency of $\omega=0.0746$ a.u.
(wavelength $\lambda=610$ nm),
is that used in the experiments of refs. \cite{French,Froggy}.
The laser pulse is turned on with a sin$^{2}(\pi t/(2\tau))$ ramp;
we treat short-pulse and long-pulse cases in which the
ramp time $\tau$ is 10 or 70 optical periods $T$ (20 or 140 fs)
respectively.

Fig. \ref{shortstrong} shows short-pulse results for a
peak field strength $F_p = 0.6$ a.u. (corresponding to a laser
intensity $I = 1.3 \times 10^{16}$ W cm$^{-2}$). The inset to Fig.
\ref{shortstrong} (a) shows the Coulomb explosion: the
nuclei remain essentially at rest until a time
just before $F_p$ is reached; then there is rapid
acceleration (increase in cycle-averaged kinetic energy) during an interval
of about 10 $T$; and then a gradual deceleration.  The dynamics
of this system can be readily understood in terms of the
time dependence of the charge distribution, which is
summarized in Fig.
\ref{shortstrong}(b). The lowest, nearly straight
line, labelled "grid" displays the total electric charge on our
grid, and thus represents the number of electrons that have been
absorbed at the boundaries; the "atomic ion" value consists of
the charge within 3 a.u. of either of the
ionic fragments, which is seen to be a good long-time
indicator of the charge localized on that fragment;
and "molecular ion" indicates the sum of the two atomic
ion charges, plus all charge in between.
We can see that by the time the molecular and atomic ions
have settled into steady behavior, at $t \approx 20 T$,
no charge has been absorbed at the walls of our box;
the gradual
departure of charge from the system at later times
does not significantly affect them.  Note that the final
charge state of each atom is +4, whereas that of the
"molecule" is +5.  Thus there are three units of
negative charge (electrons)
left in the space between the two fragments,
which screen their mutual Coulomb repulsion.

The molecular charge is seen to show a sudden decrease
at $t \approx 10T$, which corresponds to an increase
in the enclosed number of electrons.  This results
from two effects: the shutoff of molecular ionization;
and the increasing distance between the fragments, so that
more of the ejected electron cloud lies between them.  It
is not due to a rescattering mechanism, which would occur
on the time scale of 1 field period.  The mechanism of
ionization shutoff is shown in Fig. \ref{potentials},
which displays snapshots of the self-consistent molecular potential
$e\Phi(x,t)$ early in the pulse.  The
two atomic potential wells are superposed on the linear
potential of the electric field, which oscillates in time.
Electronic charge on the sides of the atomic wells or
in the interestitial region will be stripped off
early in the pulse, since there is no barrier
to its escape.  However, by $t=8.75T$ the loosely-bound
charge is gone, and the remaining charge is stuck in
the atomic wells (tunneling does not occur in our model;
it would be slow in real systems under these conditions).
The shutoff of the ionization thus stops the molecular 
charge from increasing; the actual {\it decrease} of charge
that follows is due to the enclosure of previously-ejected
electronic charge between the separating ion fragments.
Comparison of frames (a) and (b) of Fig. \ref{shortstrong}
shows that roughly two units of electronic charge are added
to the molecule in this way within about 20 fs (10 $T$) after
ionization shutoff, at which time the deceleration begins.
This results in $\Delta \approx 20$ eV, in
agreement with the experiments of ref. \cite{French,Froggy}.

For completeness we make some observations on long-pulse case.
If the ramp of the pulse
is increased, we see essentially the same
deceleration mechanism as in the short-pulse case,
although PDI occurs at high intensities.  Fig.
\ref{longstrong} shows molecular 
charges for two intensities, at the same frequency as in
Fig. \ref{shortstrong}, but with a ramp time of
140 fs.  At both intensities we see
a knee in the curve similar to that
in Fig. \ref{shortstrong}, and it occurs at about
the same value of instantaneous field strength.  However,
at the higher intensity, the knee represents only
a temporary disruption of ionization, and PDI 
occurs as the field rises to its peak strength.

In conclusion, we have presented the first dynamical model
of MEDI. The results are in qualitative agreement with experimental 
data. Reproducing the main features of the experiments, 
the model does not support the idea of stabilization \cite{French},
nor does it attribue much importance to the enhanced ionization
at some critical distance \cite{Ivanov-Corkum}.
We predict some post-dissociation
ionization for longer, picosecond pulses with peak intensities
above $10^{16}$ W cm$^{-2}$.  This could be tested experimentally.

\acknowledgements
We thank B. G. Englert for an enlightening conversation
on Thomas-Fermi methods in one and two dimensions.
This work was supported by MCS Grant No. PAN/NIST-93-156
and by KBN Grant No. 2P03B04209.

\begin{figure}
\caption{Potential energy curves for
Cl$_2$ and Cl$_2^{+2}$,
as computed with
the parameters $A=0.3, b=1.22, c=1.65$ a.u.  Inset:
Internuclear distance as a function of time for
free oscillations of Cl$_2$ for small internuclear excursions
away from $R_e$.}
\label{BOcurves}
\end{figure}

\begin{figure}
\caption{Evolution of the molecular features for
$F_p=0.6$ a.u. and $\omega=0.0746$ a.u.
with the field ramped to maximum
intensity in 10 optical periods $T$, as marked by the arrow.
(a) Kinetic energy of the atomic ion fragments vs. time;
inset shows internuclear distance vs. time.  Note
that Coulomb explosion begins near the peak intensity;
deceleration of the fragements sets in around $t=20T$.
(b) Distibution of net electrical charge (includes
both nuclear and electronic components).  Grid:
net charge of the system on our spatial grid, which
is contained in a box of length 200 a.u.; increase around
$t=20T$ comes from ejected electrons leaving the box,
as simulated by absorbing boundary conditions.
Atomic ions and molecular ions: see text.}
\label{shortstrong}
\end{figure}

\begin{figure}
\caption{Self-consistent potential curves for the case
of Fig. \protect{\ref{shortstrong}} at $t = 4.25, 7.25,$ and 8.75
$T$, as indicated.}
\label{potentials}
\end{figure}

\begin{figure}
\caption{Molecular charge vs. time
$\omega=0.0746$ a.u. in the case of a
long pulse ramp, at two values
of peak intensity as labelled: $F_p$ 
is attained at 140 fs as marked by arrow. 
The effect of ionization shutoff
and ejected-electron enclosure 
are similar to those seen in the short-pulse case
of Fig. \protect{\ref{shortstrong}}.  
The knee occurs at roughly the same
instantaneous field strength in the two cases,
corresponding to the shutoff condition displayed in
the bottom frame of Fig. \protect{\ref{potentials}}.
PDI is seen at the higher intensity}.
\label{longstrong}
\end{figure}

\end{document}